# Electrical Transport Property of ZnO Thin Films in High H$_2$ Pressure up to 20 bar

Hyunggon Chu, Byung Hoon Kim, Joonhee Kang[*]

*Department of Physics, Incheon National University, Incheon 22012*

We have investigated the H$_2$ pressure-dependent (from vacuum to 20 bar) current-voltage characteristics of ZnO thin films prepared by spin coating method. The gas pressure effect on conductance ($G$) was subtracted using He gas. The $G$ increased as applying 2 bar of H$_2$ pressure, and then it monotonously decreased with the further increment of H$_2$ pressure. Using X-ray diffraction patterns and X-ray photoelectron spectroscopy before and after H$_2$ exposure, we found that the H$_2$ spillover effect plays an important role in the variation of $G$ of ZnO film.

Email: jhkang@incheon.ac.kr

Fax: +82-32-766-8018

---

[*] Author to whom correspondence should be addressed electronic mail: jhkang@incheon.ac.kr

# I. INTRODUCTION

ZnO has a hexagonal Wurzite crystal structure with the energy band gap of 3.37 eV. Since a zinc oxide (ZnO) has the excellent properties such as *n*-type semiconductor, optical transparency in the visible light range, and a high refraction coefficient, it has attracted a great deal of interests as the materials for transparent electrode, solar cell, light emitting element, and diode [1-6]. In addition, the piezoelectricity has been extensively investigated due to its potential applications in the fields of energy harvesting (generation of voltage), detection of sound, and actuators [7-8]. Hence, the structural modification of ZnO nanowire [9], synthesis of flexible ZnO nanogenerator [10], and fabrication of flexible field-effect transistors [11] with ZnO nanowires have also been studied to enhance their piezoelectric performances.

Moreover, ZnO thin film has been focused on due to its gas sensing properties. Especially, ZnO thin film is known to be very sensitive to hydrogen gas [12]. Hydrogen sensing mechanism has been understood as follows. Oxygen molecules are adsorbed on ZnO surface when ZnO is exposed to dry air. As a result, these oxygens become ionized to form $O^-$ or $O^{2-}$ by trapping the conduction electrons from ZnO thin film. In this reaction, the electrical conductivity of *n*-type semiconductor ZnO decreases because the number of the majority carriers is reduced. On the contrary, the conductivity of ZnO thin film increases upon exposure to hydrogen gas. It results from the creation of $H_2O$ molecules due to the interaction between the ionized oxygens and hydrogen gas [13-14]. In the respect of piezoelectricity and $H_2$ sensing property, the investigation on $H_2$ pressure dependent charge transport behavior of ZnO can provide an insight of the intrinsic interaction between $H_2$ molecules and ZnO.

Here, we report that $H_2$ pressure (vacuum ~ 20 bar) -dependent conductance of ZnO thin film prepared by the sol-gel and spin coating methods [15] to investigate on the interaction with $H_2$ molecules and the piezoelectric phenomena, simultaneously. The piezoelectric effect was obtained from the conductance as a function of helium pressure from vacuum to 20 bar. The structural change of ZnO after $H_2$ exposure was confirmed by X-ray diffraction patterns (XRD) and X-ray photoelectron spectroscopy (XPS) before and after $H_2$ exposure. Both the charge transport properties and structural modulation were interpreted by hydrogen spillover effect.

## II. EXPERIMENTS

We synthesized the ZnO films with zinc acetate dihydrate ($Zn(CH_3COO)_2 \cdot 2H_2O$, $Zn(OAc)_2$), 2-methoxyethanol (2MOE), and monoethanolamine (MEA) [15]. $Zn(OAc)_2$ was dissolved in 2MOE with MEA as stabilizer. The silicon substrates were washed with acetone, methanol, and deionized water (DI water). ZnO films were prepared on the substrates by spin coating. The spin coating speed was 1000 rpm for 30 sec. The prepared films were baked on hotplate at 200 ºC for 10 min, followed by sintering in the tube furnace. This procedure was repeated three times. The temperature in the tube furnace was increased from room temperature to 700 ºC at the speed of 20 ºC /min, and kept at 700 ºC for 1 hr, then cooled down to room temperature naturally.

X-ray diffraction (XRD, SmartLab / Rigaku) was used to examine the crystal structure of the ZnO thin films. The chemical states of ZnO films was investigated by X-ray Photoelectron Spectroscopy (XPS, PHI 5000 Versa Probe Ⅱ). The conventional four-probe conductances of the bar-type ZnO thin films were measured at 300 K and in the high pressure He and $H_2$ (99.999 %) atmosphere, using 4200-SCS semiconductor characterization system (Keithley).

## III. RESULTS AND DISCUSSION

Figure 1 shows the current-voltage (*I-V*) characteristics of ZnO films as a function of He (Fig. 1(a)) and $H_2$ (Fig. 1(b)) gas pressure from vacuum (~$10^{-6}$ Torr) to 20 bar. First, *I-V* curves were obtained with He gas, and then the evacuation process was performed at 373 K to remove remnant He gas. Finally, $H_2$ pressure-dependent *I-V* curves were obtained with the same ZnO film. Upon exposure of ZnO films to 2 bar of He and $H_2$ gases, the slope in the *I-V* curve abruptly increased compared with that in a vacuum. The change of the slope became complex as the pressure increased. This behavior is easily verified with the gas pressure-dependent conductance (*G*) as shown in Fig. 1(c). As soon as the ZnO films were exposed to 2 bar of gas pressure, we observed the increase of *G* in the both cases ($H_2$ and He). However, *G* decreased up to 6 bar and then it increased gradually from 8 to 20 bar. It is expected that the variation of *G* affected only by pressure is extracted from the He-pressure dependent *G* because He is an inert gas. On the whole, *G* increases with the increase of He pressure, which comes from the piezoelectric potential induced by a strain. However, the *G* decreased from 2 to 6 bar of He pressure as mentioned above. We suggest that the reduction of *G* can be explained by the competition of polarization directions because the piezoelectric potential of ZnO depends on a bending directions; the current induced by piezoelectricity increases (decreases) as compressive (stretching) force is applied to the ZnO wire [9]. Although the pressure is applied to all directions of ZnO, the amount of change of each bond length such as Zn-O and Zn-Zn is different [16]. It causes the variation of coordination [17] and the modulation of energy bandgap [18]. This anisotropic modulation of the structure can be the reason for the competition of polarization directions. When we exposed ZnO film to $H_2$ gas, the large *G* was observed compared with that in He gas, a result from $H_2$ sensing property of ZnO. To show the *G* variation without piezoelectric effect, we subtracted the *G* for He pressure from *G* for $H_2$ pressure (Fig. 1(d)). The significant increase of conductance was observed at 2 bar. However, the *G* decreased monotonously as $H_2$ pressure increased. On the contrary to He exposure, the conductance did not recover the original value even after exposed to the high vacuum ($10^{-6}$ Torr) at 373 K (green diamond in Fig. 1(c)). It means that the structural modulation occurs due to $H_2$ exposure. To investigate the structural change, XRD patterns were obtained before and after $H_2$ exposure.

Figure 2 shows the XRD patterns before and after $H_2$ exposure. The characteristic peaks for ZnO film was well defined before $H_2$ exposure. The peaks shifted toward lower angles after the exposure of ZnO film to high $H_2$ pressure. Figure 2(b) depicts this behavior with three main peaks, indicating that the lattice space of ZnO becomes slightly larger due to $H_2$ exposure. Moreover, new peaks near $2\theta$ = 38.24 and 44.23° were developed (diamond in Fig. 2(a)), which correspond to the XRD patterns for $Zn(OH)_2$ peaks (JCPDS 38-0385) [19]. It means that the chemical states of ZnO change due to the interaction with $H_2$ molecules.

In order to find the change of chemical states of ZnO, we performed an XPS study of the ZnO film before (left panel) and after $H_2$ exposure (right panel) as shown in Fig. 3. Zn-O bonding at 530.3 eV, hydroxyl groups at 531.4 eV, and molecular water at 532.1 eV were observed in O 1s spectra, which are well consistent with the previous report [20-21]. The variations of the amount of these three species before and after $H_2$ exposure were compared (Fig. 3(a)). The amount of Zn-O decreased from 46.74 to 35.10 %, but that of hydroxyl groups (from 3.12 to 5.92 %) and molecular water (from 50.14 to 58.98 %) increased after the reaction with $H_2$ molecules. Figure 3(b) shows the Zn 2p spectra. P1 (1021.5 eV) and P2 (1022.2 eV) peaks correspond to Zn-O and $Zn(OH)_2$, respectively. The area of P1 decreased but that of P2 increased due to $H_2$ exposure. This indicates that the increment of OH species (from 37.88 to 44.53 %) occurs.

From the results obtained from the electrical transport properties, XRD, and XPS, we propose the spillover phenomenon of $H_2$ gas molecules on ZnO. First, Hydrogen molecules are catalytically dissociated on ZnO (step I). Simultaneously, some ZnO species are reduced and molecular water are produced as demonstrated by XPS study. Second, dissociated hydrogen atoms which do not participate in the reaction of step I migrate to the ZnO surface and diffuse into ZnO (step II). Finally, these hydrogen atoms break the Zn-O bonds. It results in the production of delocalized electrons which cause to increase in conductance. Moreover, OH (hydroxyl) groups and H-O-H bonds are created (step III). Consequently, the structure of ZnO is modulated, causing that the conductance does not increase any more as shown in Fig. 1(d).

## IV. CONCLUSIONS

In summary, we report the structural change and electrical transport properties of ZnO thin film in high $H_2$ gas pressure up to 20 bar. The piezoelectricity of ZnO film was investigated using high He pressure, and this piezoelectric effect was subtracted to show the sole $H_2$ pressure effect on $G$ of ZnO. We found that the $G$ increased when ZnO film was exposed to 2 bar of $H_2$ and then it gradually decreased with the further increment of $H_2$ pressure. From structural and chemical state modulation confirmed by XRD and XPS spectra O 1s and Zn 2p before and after $H_2$ exposure, we suggest that the spillover of $H_2$ molecules produces $Zn(OH)_2$ species. The competition between delocalized electrons and the structural modulation originated from the spillover process results in the $H_2$ pressure-dependent $G$ of ZnO film.

## ACKNOWLEDGEMENT

The authors would like to thank Prof. Lili Yang in Jilin Normal University for the early discussions in sample preparation. This work was supported by the Incheon National University Research Grant in 2013 (20130537).

**Figure Captions**

Figure 1. (a) He and (b) $H_2$ pressure-dependent *I-V* characteristics of ZnO film from vacuum to 20 bar.

(c) Pressure-dependent $G$ obtained from $I$-$V$ curves. The green diamond indicates the conductance after $H_2$ exposure followed by the evacuation process. (d) He pressure-dependent $G$ was subtracted from $H_2$ pressure-dependent $G$ to exclude the pressure effect.

Figure 2. (a) XRD patterns of ZnO thin film before and after $H_2$ exposure showed that the ZnO films were well synthesized and then the new two peaks for $Zn(OH)_2$ were developed after $H_2$ exposure (red diamond). (b) The peaks for ZnO shifted to lower angles, indicating lattice spacing became larger due to hydrogen spillover.

Figure 3. XPS spectra of (a) O 1s and (b) Zn 2p before (left panel) and after (right panel) $H_2$ exposure. The results show that ZnO species are slightly suppressed, but the species of hydroxyl and molecular water increases. P1 and P2 in (b) represent for Zn-O and hydroxyl groups, respectively.

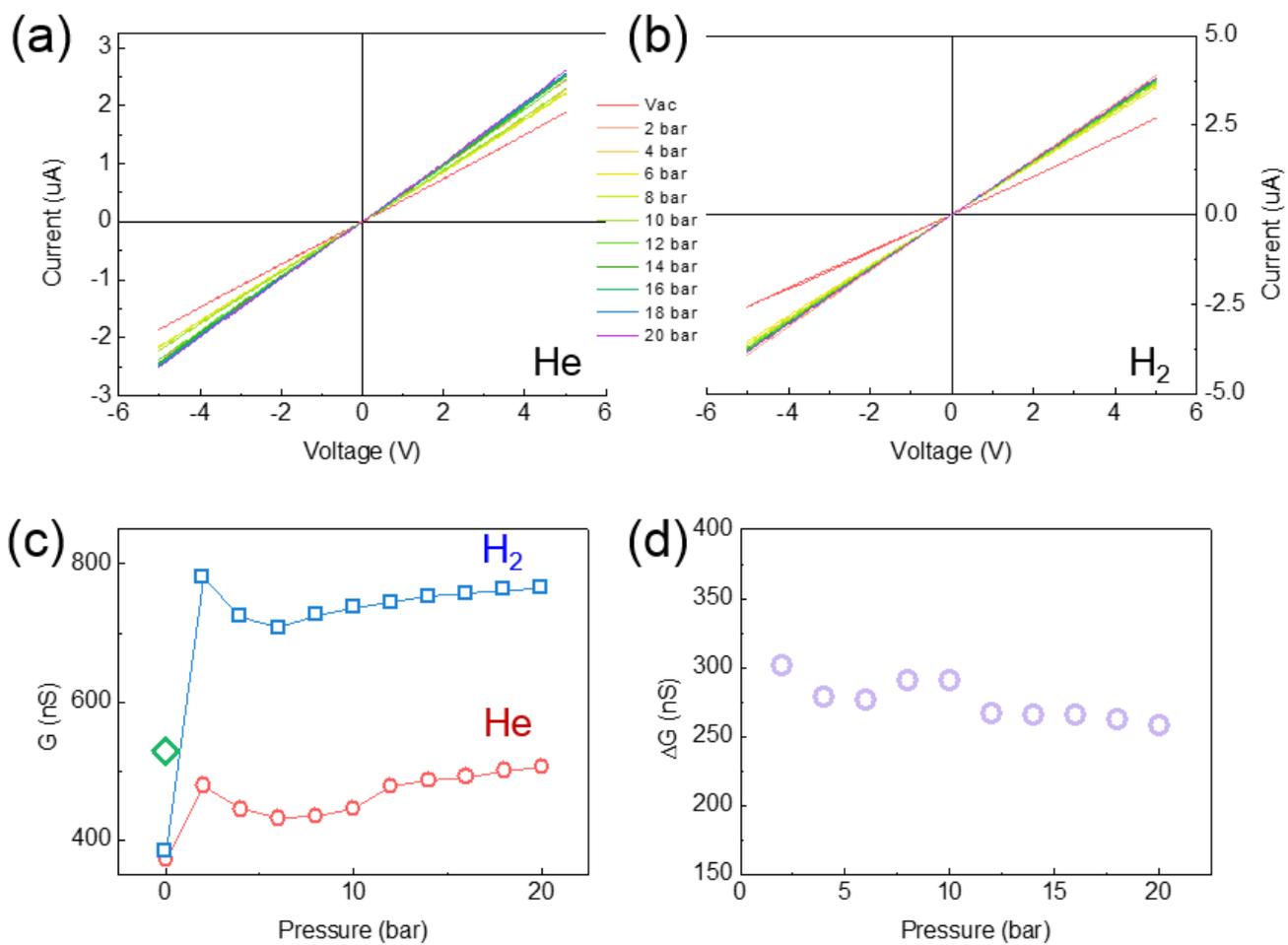

Fig. 1.

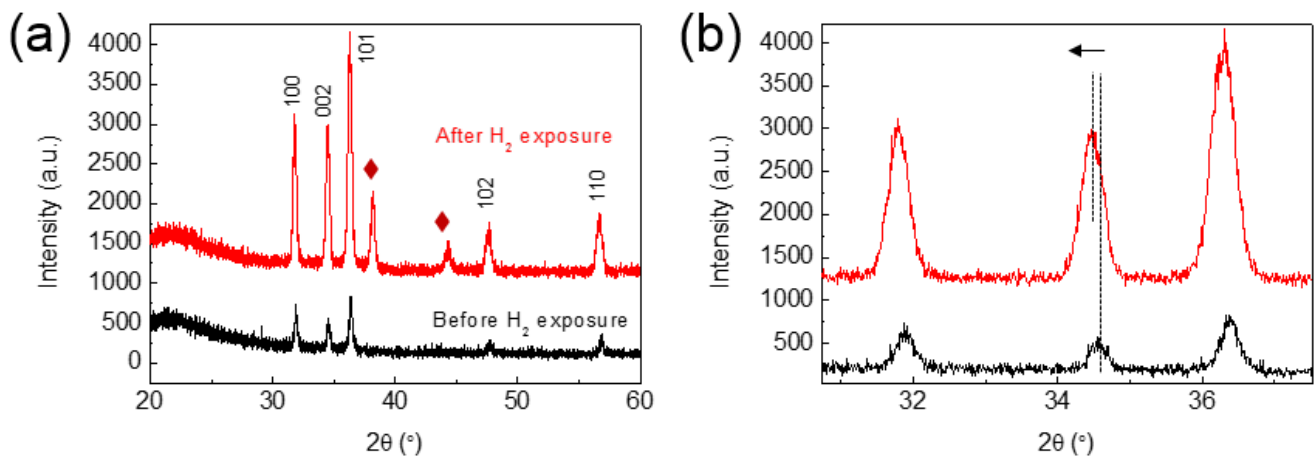

Fig. 2

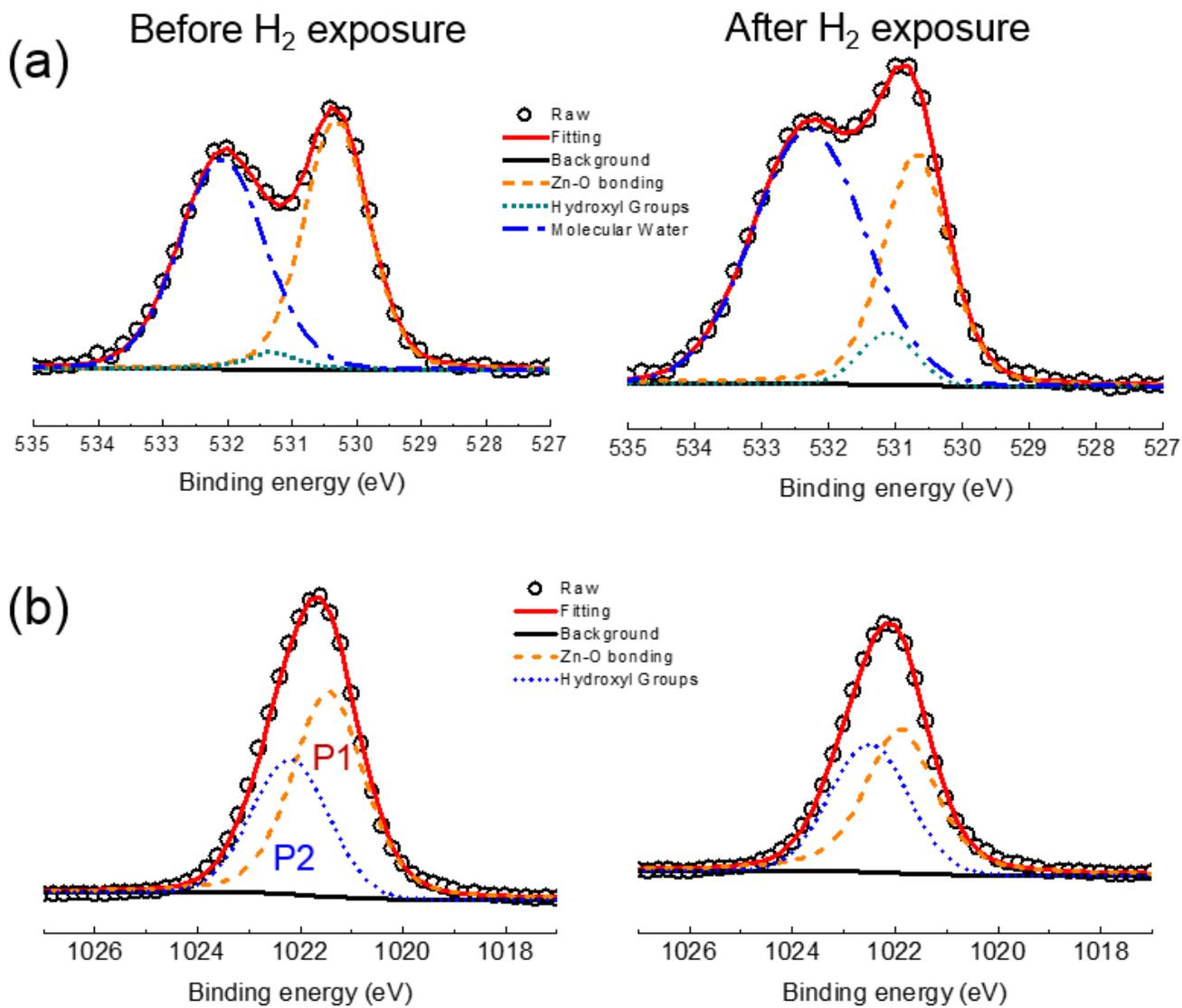

Fig. 3